\documentclass{jfm}
\usepackage{graphicx}
\usepackage{epstopdf, epsfig}
\usepackage{amsmath}
\usepackage{xcolor}

\shorttitle{Rotation of bulk turbulent convection}
\shortauthor{F. Toselli, S. Musacchio and G. Boffetta}

\title{Effects of rotation on the bulk turbulent convection}

\author{Francesco Toselli\aff{1}
  \corresp{\email{francesco.toselli@unito.it}},
  Stefano Musacchio\aff{1}
 \and Guido Boffetta\aff{1}}

\affiliation{\aff{1}Dipartimento di Fisica and INFN, Universit\`a di Torino, 
via P. Giuria 1, 10125 Torino, Italy
\aff{2}}

\begin{document}

\maketitle

\begin{abstract}
We study rotating homogeneous turbulent convection forced by a mean
vertical temperature gradient by means of direct numerical simulations 
(DNS) in the Boussinesq approximation in a rotating frame.
In the absence of rotationour results are in agreement with the ``ultimate regime 
of thermal convection'' for the scaling of the Nusselt and Reynolds numbers
vs Rayleigh and Prandtl numbers. 
Rotation is found to increase both $Nu$ and $Re$ at fixed 
$Ra$ with a maximum enhancement for intermediate values of 
the Rossby numbers, qualitatively similar, but with stronger intensity,
to what observed in Rayleigh-B\'enard rotating convection. 
Our results are interpreted in terms of a quasi-bidimensionalization of the 
flow with the formation of columnar structures displaying strong 
correlation between the temperature and the vertical velocity fields.
\end{abstract}

\begin{keywords}
direct numerical simulation, rotating convection, bulk turbulent convection
\end{keywords}

%%%%%%%%%%%%%%%%%%%%%%%%%%%%%%%%%%%%%%%%%%%%%%%%%%%%%%%%%%%%%%%%%
\section{Introduction} 
\label{sec1}
Turbulent convection involves the coupling between an active temperature
field transported by a a turbulent flow in presence of gravity. 
Within this general framework, different examples of turbulent convection
are characterized essentially by boundary conditions which 
force the flow in different ways. 
In the most common configurations temperature difference
is parallel to gravity, as in the case of Rayleigh-B\'enard 
(RB) convection, in which the flow is confined into a box with fixed 
temperatures on the two horizontal boundaries 
\citep{Bodenschatz2000,Ahlers2009} or for Rayleigh-Taylor (RT) convection, 
which is forced by two reservoirs of fluid at different temperature
\citep{boffetta2017incompressible}.
Another geometry, which has become very popular for numerical simulations,
is the so-called {\it bulk turbulent convection} (BTC) in which the flow 
is forced by an imposed vertical temperature linear gradient. 
BTC is motivated by the study of the ultimate state regime predicted 
by \citet{Kraichnan1962}, which is supposed to appear
in RB convection when the contribution of boundary layers become negligible
\citep{grossmann2004fluctuations}. 
Moreover, it is similar to the turbulent phase of RT convection where a 
linear temperature (density) profile naturally appears and
both RT and BTC display the ultimate state regime
\citep{lohse2003ultimate,calzavarini2005rayleigh,boffetta2012ultimate}.

Several internal and external factors can modify the dynamical and
statistical properties of turbulent convection: 
among the latters, rotation along the
vertical axis is known to affect 
the efficiency of turbulent transport of heat in both RB and RT convection.
The study of the effects of rotation is of great interest 
because of its relevance for geophysical and astrophysical applications,
including
convection in the oceans \citep{Marshall1999} and in the 
atmosphere \citep{hartmann2001tropical,Rahmstorf2000}), convection
inside gaseous giant planets \citep{Busse1994} or in external layer of the Sun
\citep{Miesch2000}), and for technological applications \citep{Johnston1998}.

Linear stability analysis, performed originally by \citet{Chandrasekhar1961}
for RB shows that rotation has a stabilizing effect 
and this suggests that it might reduce the transfer of heat in the 
nonlinear, turbulent phase. 
However, the work by \citet{Rossby1969} shows that rotation can also 
increase the heat transport. This enhancement is explained by the 
mechanism of Ekman pumping \citep{Zhong2009,King2009,Kunnen2008,Julien1996} 
that contributes to a vertical heat flux produced by an extra vertical 
circulation due to a suction of fluid at the two boundary layers.
The effect of rotation in turbulent RB convection has been extensively 
studied by means of experiments 
\citep{Brown2005,Kunnen2008,Niemela2010,Kunnen2011} 
and numerical simulations \citep{Sprague2006,Stevens2009,Stevens2010,Chong2017}.
The picture which emerges is that the heat transport between the hot and 
the cold plate, measured by the dimensionless Nusselt number $Nu$ 
(all parameters are defined below), has a non-monotonic dependence on
the rotation, identified by the dimensionless Rossby number $Ro$: 
moderate rotations enhance the heat transfer while stronger rotations 
bring to an important suppression of the vertical velocities and to a 
reduction of the heat transport. 

In the case of RT convection, the effect of rotation has been studied 
more recently by means of both experiments \citep{baldwin2015inhibition}
and DNS within the
Boussinesq approximation \citep{boffetta2016rotating}.
The main result is that rotation always reduces the turbulent heat
transfer in this case. 
The mechanism for this reduction is due to a partial decoupling 
and decorrelation of the temperature and the vertical velocity fields 
which reduces the Nusselt number.
This result does not contrast with the enhancing mechanism 
associated to the Ekman pumping which has been observed
in the RB case, because of the absence of boundary layers 
in the RT system. 

The aim of this paper is to investigate the effects of rotation on the
heat transfer within the framework of 
the BTC, driven by a mean temperature gradient. 
Surprisingly, at variance with RT convection, we find a
strong enhancement of the Nusselt number (at fixed Rayleigh
number) induced by rotation. A detailed analysis shows that
the heat flux is mainly due to the formation of convective columnar 
structures produced by the quasi-bidimensionalization of the flow.

The remaining of this paper is organized as follow. Section~\ref{sec2} is
devoted to the description of the numerical simulations while
in section~\ref{sec3} we discuss the dependence 
of Nusselt and Reynolds number on rotation. In Section
\ref{sec4} we investigate the role played by the columnar structures
generated by the rotation in the process of heat transfer.  
Finally, conclusions are reported in Section~\ref{sec5}. 

%%%%%%%%%%%%%%%%%%%%%%%%%%%%%%%%%%%%%%%%%%%%%%%%%%%%%%%%%%%%%%%%%
\section{Mathematical model and numerical method}
\label{sec2}
We perform extensive numerical simulations of BTC
by integrating the Boussinesq equations for an incompressible 
flow forced by a mean unstable temperature gradient $-\gamma$
in a cubic box of size $L$ \citep{borue1997turbulent,lohse2003ultimate}.
The temperature field is therefore written as 
$T({\bf x},t)=-\gamma z + \theta({\bf x},t)$, where $\theta({\bf x},t)$
represents the fluctuation field.
The Boussinesq equations, written in a reference frame rotating with
angular velocity ${\bf \Omega}=(0,0,\Omega)$ along the $z$ axis, read
\begin{equation}
\partial_t {\bf u} + {\bf u} \cdot {\bf \nabla} {\bf u} 
+ 2 {\boldsymbol \Omega} \times {\bf u}
= - {\bf \nabla} p + \nu \nabla^2 {\bf u} - \beta {\bf g} \theta
\label{eq1}
\end{equation}
\begin{equation}
\partial_t \theta + {\bf u} \cdot {\bf \nabla} \theta = \kappa \nabla^2 \theta +
\gamma w
\label{eq2}
\end{equation}
where ${\bf u}=(u,v,w)$ is the incompressible (${\bf \nabla}\cdot{\bf u}=0$)
velocity field, $p$ is the pressure, $\beta$ is the thermal expansion coefficient, 
${\bf g}=(0,0,-g)$ is gravity, $\nu$ 
is the kinematic viscosity and $\kappa$ the thermal diffusivity.  

The dimensionless parameters which govern the flow are the Rayleigh
number, defined as $Ra=\beta g \gamma L^4/(\nu \kappa)$ 
(where $L$ is the size of the system), the Prandtl number 
$Pr=\nu/\kappa$ and the Rossby number, here defined as 
$Ro=\sqrt{\beta g \gamma}/(2 \Omega)$, which measures the (inverse)
intensity of rotation as the ratio between the buoyancy and Coriolis force.
When the turbulent flow reaches a statistical stationary condition, 
we measure velocity and temperature fluctuations and their correlation
from which we compute the Reynolds number 
$Re=UL/\nu$ (where $U=\sqrt{\langle\mid{\bf u}^{2}\mid\rangle / 3}$ is
the root mean square of all velocity components) 
and the Nusselt number is defined as 
$Nu=\langle w \theta \rangle/(\kappa \gamma)+1$ 
with $\langle...\rangle$ indicating the average over the volume. 
%We note that the mean temperature gradient does not contribute to
%$Nu$, since the average of $w$ on horizontal planes vanishes.

%%%%%%%%%%%%%%%%%%%%%%%%%%%%%%%%%%%%%%%%%%%%%%%%%%%%%%%%%%%%%%%%%%%%%%%%%%
\begin{table}
\begin{center}
\begin{tabular}{crccclcc}
\multicolumn{1}{c}{Ra} & \multicolumn{1}{c}{Pr} &
\multicolumn{1}{c}{Ro} & \multicolumn{1}{c}{Nu} & \multicolumn{1}{c}{Re} & \multicolumn{1}{c}{$\Omega$} & \multicolumn{1}{c}{$\nu$} & \multicolumn{1}{c}{$\kappa$} \\[4pt]
$1.1 \times 10^7$     & 10 & $\infty$     				    &$3.12 \times 10^{3}$&$4.57 \times 10^{2}$& $0$ &  $6.00 \times 10^{-3}$  & $6.00 \times 10^{-4}$  \\
$1.1 \times 10^7$     & 10 & $3.16 \times 10^{-1}$ 	&$5.86 \times 10^{3}$&$6.04 \times 10^{2}$& $0.25$ &  $6.00 \times 10^{-3}$  & $6.00 \times 10^{-4}$  \\
$1.1 \times 10^7$     & 10 & $1.58 \times 10^{-1}$ 	&$8.19 \times 10^{3}$&$6.99 \times 10^{2}$& $0.5$ &  $6.00 \times 10^{-3}$  & $6.00 \times 10^{-4}$  \\
$1.1 \times 10^7$     & 10 & $7.91 \times 10^{-2}$  &$1.14 \times 10^{4}$&$8.12 \times 10^{2}$& $1$ &  $6.00 \times 10^{-3}$  & $6.00 \times 10^{-4}$  \\
$1.1 \times 10^7$     & 10 & $3.95 \times 10^{-2}$  &$9.56 \times 10^{3}$&$7.60 \times 10^{2}$& $2$ &  $6.00 \times 10^{-3}$  & $6.00 \times 10^{-4}$  \\
$1.1 \times 10^7$     & 10 & $1.98 \times 10^{-2}$  &$9.02 \times 10^{3}$&$7.38 \times 10^{2}$& $4$ &  $6.00 \times 10^{-3}$  & $6.00 \times 10^{-4}$  \\
$2.2 \times 10^7$     & 1   & $\infty$                                  &$1.97 \times 10^{3}$&$2.48 \times 10^{3}$& $0$ &  $1.89 \times 10^{-3}$  & $1.89 \times 10^{-3}$  \\
$2.2 \times 10^7$     & 1   & $4.47 \times 10^{-1}$ 	&$2.92 \times 10^{3}$&$2.94 \times 10^{3}$& $0.25$ &  $1.89 \times 10^{-3}$  & $1.89 \times 10^{-3}$  \\
$2.2 \times 10^7$     & 1   & $2.23 \times 10^{-1}$ 	&$3.87 \times 10^{3}$&$3.31 \times 10^{3}$& $0.5$ &  $1.89 \times 10^{-3}$  & $1.89 \times 10^{-3}$  \\
$2.2 \times 10^7$     & 1   & $1.12 \times 10^{-1}$ 	&$5.18 \times 10^{3}$&$3.77 \times 10^{3}$& $1$ & $1.89 \times 10^{-3}$   & $1.89 \times 10^{-3}$  \\
$2.2 \times 10^7$     & 1   & $5.59 \times 10^{-2}$  &$5.29 \times 10^{3}$&$3.84 \times 10^{3}$& $2$ &  $1.89 \times 10^{-3}$  & $1.89 \times 10^{-3}$  \\
$2.2 \times 10^7$     & 1   & $2.79 \times 10^{-2}$  &$3.43 \times 10^{3}$&$3.41 \times 10^{3}$& $4$ &  $1.89 \times 10^{-3}$  & $1.89 \times 10^{-3}$  \\
$2.2 \times 10^7$     & 5   & $\infty$	 			         &$3.67 \times 10^{3}$&$1.02 \times 10^{3}$& $0$  &  $4.24 \times 10^{-3}$  & $0.85 \times 10^{-3}$  \\
$2.2 \times 10^7$     & 5   & $4.47 \times 10^{-1}$ 	&$4.86 \times 10^{3}$&$1.15 \times 10^{3}$& $0.25$ &  $4.24 \times 10^{-3}$  & $0.85 \times 10^{-3}$  \\
$2.2 \times 10^7$     & 5   & $2.23 \times 10^{-1}$ 	&$7.70 \times 10^{3}$&$1.41 \times 10^{3}$& $0.5$ &  $4.24 \times 10^{-3}$  & $0.85 \times 10^{-3}$  \\
$2.2 \times 10^7$     & 5   & $1.12 \times 10^{-1}$ 	&$1.10 \times 10^{4}$&$1.64 \times 10^{3}$& $1$ &  $4.24 \times 10^{-3}$  & $0.85 \times 10^{-3}$  \\
$2.2 \times 10^7$     & 5   & $5.59 \times 10^{-2}$  &$1.27 \times 10^{4}$&$1.77 \times 10^{3}$& $2$ &  $4.24 \times 10^{-3}$  & $0.85 \times 10^{-3}$  \\
$2.2 \times 10^7$     & 5   & $2.79 \times 10^{-2}$  &$8.30 \times 10^{3}$&$1.59 \times 10^{3}$& $4$ &  $4.24 \times 10^{-3}$  & $0.85 \times 10^{-3}$  \\
$2.2 \times 10^7$     & 10 & $\infty$     				    & $4.88 \times 10^{3}$&$6.87 \times 10^{2}$& $0$ &  $6.00 \times 10^{-3}$  & $6.00 \times 10^{-4}$  \\
$2.2 \times 10^7$     & 10 & $4.47 \times 10^{-1}$ 	& $6.50 \times 10^{3}$&$7.78 \times 10^{2}$& $0.25$ &  $6.00 \times 10^{-3}$  & $6.00 \times 10^{-4}$  \\
$2.2 \times 10^7$     & 10 & $2.23 \times 10^{-1}$ 	&$9.55 \times 10^{3}$&$9.29 \times 10^{2}$& $0.5$ &  $6.00 \times 10^{-3}$  & $6.00 \times 10^{-4}$  \\
$2.2 \times 10^7$     & 10 & $1.12 \times 10^{-1}$ 	&$1.40 \times 10^{4}$&$1.09 \times 10^{3}$& $1$ &  $6.00 \times 10^{-3}$  & $6.00 \times 10^{-4}$  \\
$2.2 \times 10^7$     & 10 & $5.59 \times 10^{-2}$  &$1.60 \times 10^{4}$&$1.18 \times 10^{3}$& $2$ &  $6.00 \times 10^{-3}$  & $6.00 \times 10^{-4}$  \\
$2.2 \times 10^7$     & 10 & $2.79 \times 10^{-2}$  &$1.34 \times 10^{4}$&$1.13 \times 10^{3}$& $4$ &  $6.00 \times 10^{-3}$  & $6.00 \times 10^{-4}$  \\
\end{tabular}
\caption{Parameters of the numerical simulations}
  \label{tab:1}
  \end{center}
\end{table}
%%%%%%%%%%%%%%%%%%%%%%%%%%%%%%%%%%%%%%%%%%%%%%%%%%%%%%%%%%%%%%%%%%%%%%%%%%

We performed extensive direct numerical simulations of equations 
(\ref{eq1}-\ref{eq2}) by means of a fully parallel pseudo-spectral
code at resolution $N^3=512^3$ in a cubic domain of size $L=2\pi$ with
periodic boundary conditions. We explore the set of parameters by
considering two different Rayleigh numbers, $Ra=1.1 \times 10^7$ and 
$Ra=2.2 \times 10^7$, three values of the Prandtl number 
$Pr=1$, $Pr=5$ and $Pr=10$ and $6$ different Rossby numbers. 
The different $Pr$ numbers are obtained by changing both $\nu$ 
and $\kappa$ by keeping their product constant which fixes
the value of $Ra$. The two different $Ra$ are obtained by changing
the mean temperature gradient $\gamma$. 
All parameter values for the simulations are showed in
Table~\ref{tab:1}.
The maximum value of $Ra$ has been chosen such that 
in the case $Pr=1$ and $Ro=\infty$ both the Kolmogorov scale
$\eta=(\nu^3/\varepsilon)^{1/4}$ 
and the Batchelor scale 
$\ell_B=(\kappa^2\nu/\varepsilon)^{1/4}$
(where $\varepsilon=\nu\langle(\partial_iu_j)^2\rangle$ 
is the volume averaged kinetic dissipation rate) 
are well resolved. 
In terms of the maximum wavenumber $K_{max}=N/3$ 
we have $K_{max} \eta = K_{max} \ell_B = 2.4$ 
for the case $Pr=1$ and $Ro=\infty$.
The effects of rotation on the Kolmogorov and Batchelor scales could
not be predicted a priori, 
but we have checked a posteriori that in the worst case 
we have $K_{max} \eta > 1.8$ (for $\Omega=4$, $Pr=1$) 
and $K_{max} \ell_B > 1.4$ (for $\Omega=4$, $Pr=10$).
The duration of each simulation is $T=100\tau$, measured in units of
the characteristic time $\tau=1/\sqrt{\beta\,g\,\gamma}$. 

We found that average quantities such as $Re$ and $Nu$ display strong 
fluctuations in the time series. 
 
Therefore as a measure of the error on the time average of these
quantities we use the maximum fluctuation of the running average computed 
on the second half of the time series. 

%%%%%%%%%%%%%%%%%%%%%%
\begin{figure}
\centerline{\includegraphics[width=0.5\columnwidth]{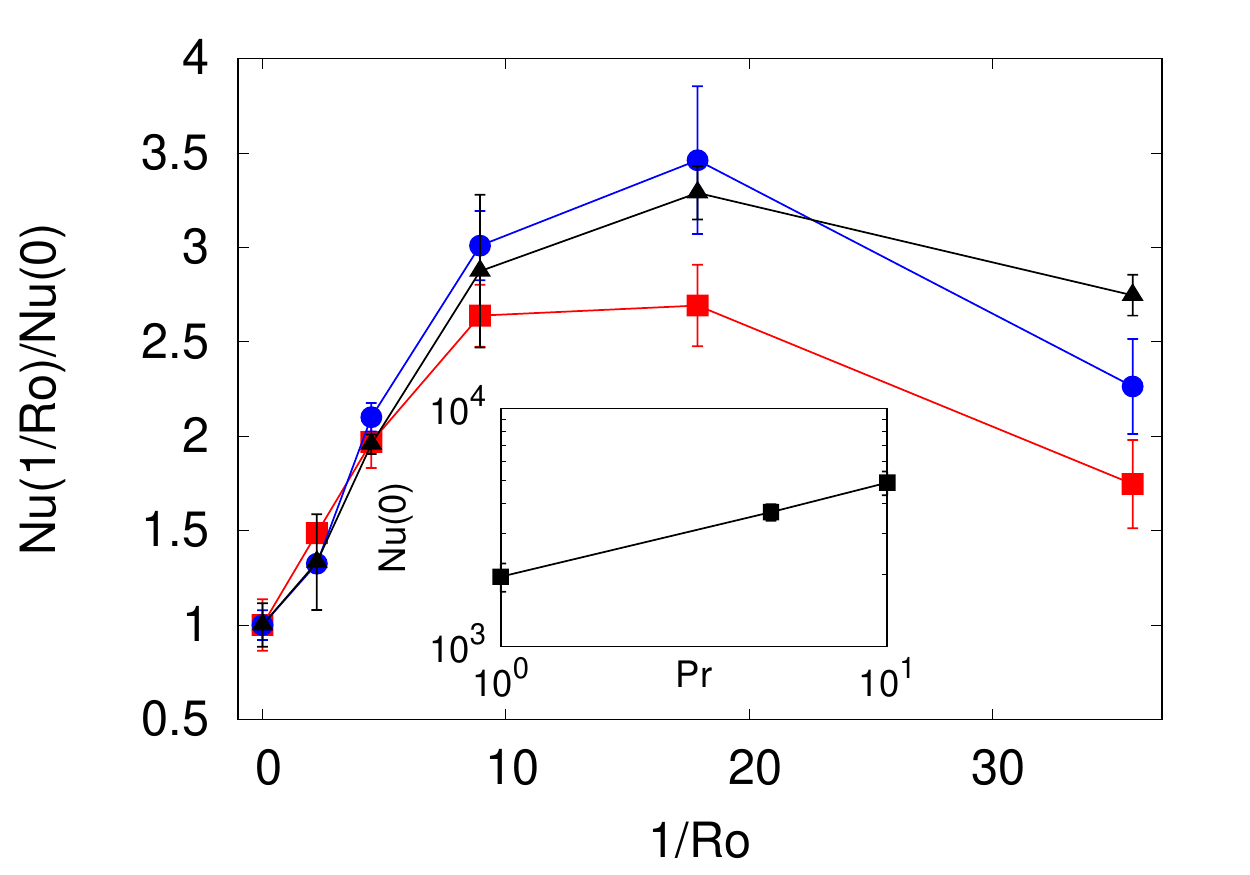}
\includegraphics[width=0.5\columnwidth]{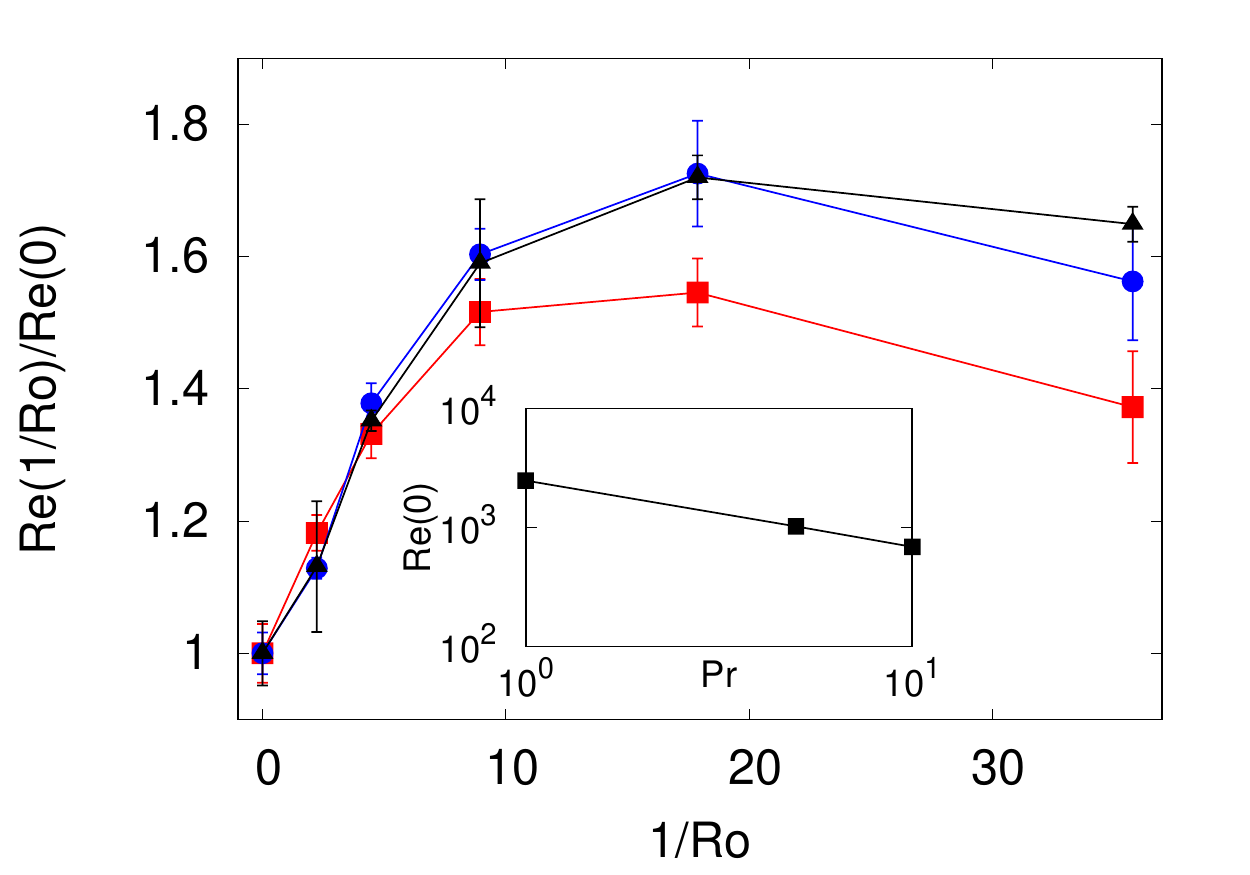}}
\caption{$Nu$ (a) and $Re$ (b) as a function
of $1/Ro$ normalized with the value at $1/Ro=0$ for simulations at 
$Ra=2.2 \times 10^7$ and 
$Pr=1$ (red squares), $Pr=5$ (blue circles) and $Pr=10$ 
(black triangles). The insets show the values of $Nu$ and $Re$ 
in the absence of rotation ($1/Ro=0$) as a function of $Pr$. 
The lines represent the scaling $Nu(0) \propto Pr^{0.40}$ and
$Re(0) \propto Pr^{-0.55}$.}
\label{fig1}
\end{figure}
%%%%%%%%%%%%%%%%%%%%%%
%%%%%%%%%%%%%%%%%%%%%%
\begin{figure}
\centerline{\includegraphics[width=0.5\columnwidth]{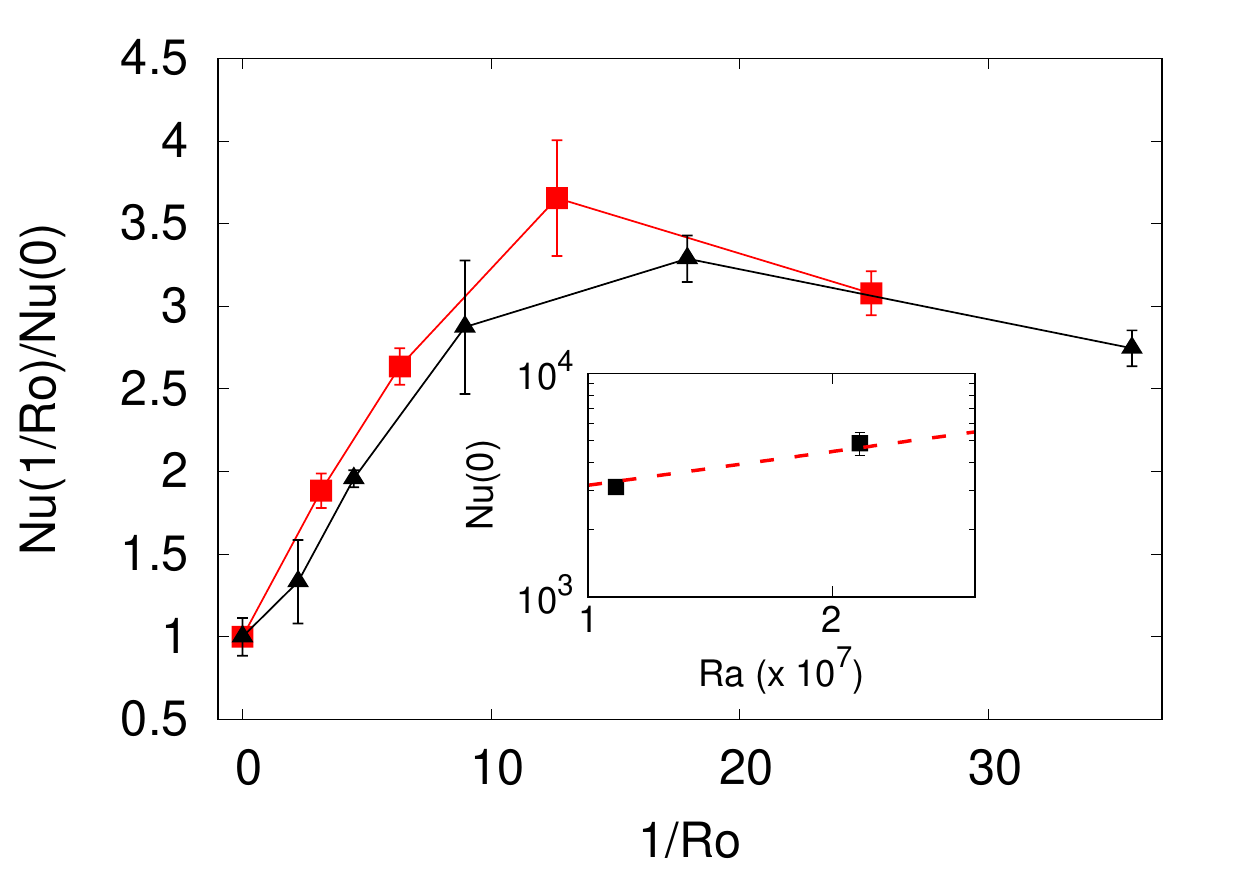}
\includegraphics[width=0.5\columnwidth]{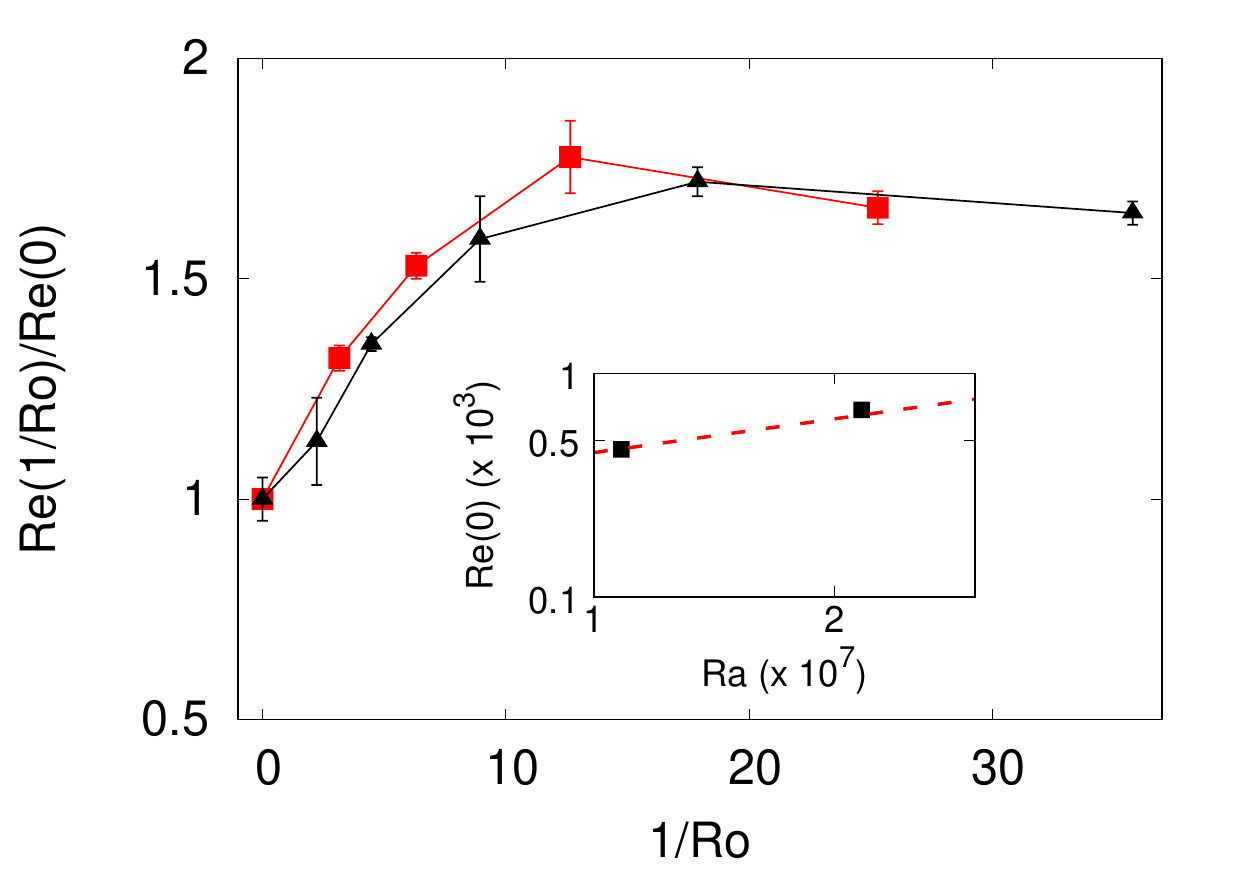}}
\caption{$Nu$ (a) and $Re$ (b) as a function
of $1/Ro$ normalized with the value at $1/Ro=0$ for simulations at 
$Ra=1.1 \times 10^7$ (red squares) and $Ra=2.2 \times 10^7$ 
(black triangles) for the case $Pr=10$. 
The insets show the values of $Nu$ ($Re$) 
in the absence of rotation ($1/Ro=0$) as a function of $Ra$. 
The dashed red lines represent the scaling $Nu(0) \propto Ra^{1/2}$ and
$Re(0) \propto Ra^{1/2}$.}
\label{fig2}
\end{figure}
%%%%%%%%%%%%%%%%%%%%%%
%%%%%%%%%%%%%%%%%%%%%%%%%%%%%%%%%%%%%%%%%%%%%%%%%%%%%%%%%%%%%%%%%
\section{Nusselt and Reynolds dependence on rotation}
\label{sec3}

In order to study the effects of the Coriolis force on the heat transfer and
the turbulence intensity, we first consider the dependence of $Nu$ and $Re$
on the rotation number $1/Ro$ for different values of $Pr$. 
In Fig.~\ref{fig1} we report the values of $Nu$ and $Re$ rescaled on
their respective value in absence of rotation ($1/Ro=0$) 
for the simulations at $Re=2.2 \times 10^7$. 
We find a non-monotonic dependence: 
the heat transfer (measured by $Nu$) and the turbulence
intensity (quantified by $Re$) increase with
the rotation rate and they attain a maximum for an optimal value of $Ro \approx 6 \times 10^{-2}$. 
For stronger rotation rates they decrease slowly. 
The relative variation with respect to the non-rotating case
($Nu(0)$) is larger for the cases $Pr=5$ and $Pr=10$.

The non-monotonic behavior of $Nu$ and $Re$ as a function of $Ro$, 
as well as the dependence on $Pr$, is qualitatively similar to what has been reported in
previous works for the case of turbulent RB convection~\citep{Zhong2009,Stevens2010Prandtl,Stevens2011,stevens2013heat}. 
The main difference between the RB case is the magnitude of the heat transfer
enhancement: in our simulations of BTC we observe a maximum relative increase
of $Nu$ of a factor $3.5$. This enhancement is much larger than the increase of 
a factor $1.1-1.2$ which has been observed 
in the RB case for $Ra$ in the range of $10^8-10^9$ \citep{stevens2013heat}. 
Moreover, the decay at large rotation rates is much
slower in BTC case than in RB case. 
It is worth to notice that the mechanisms which originate the heat transfer
enhancement are different in RB and BTC: 
in the case of the RB convection, the increase of $Nu$
is mostly due to the effects of the rotation on the boundary layers. 
The latters are absent on the BTC case, which is dominated by bulk
effects. 

In absence of rotation, the scaling of $Nu(0)$ and $Re(0)$ as a function of
$Pr$ observed in our simulations are $Nu(0) \propto Pr^{0.40}$ and $Re(0)
\propto Pr^{-0.55}$ 
(see inset of Fig.~\ref{fig1})
The scaling exponents are close to those 
predicted for the ultimate state of turbulent convection $Nu \propto Ra^{1/2}Pr^{1/2}$
 and $Re \propto Ra^{1/2}Pr^{-1/2}$ \citep{Kraichnan1962} 
and they are in agreement with previous numerical results for RB case
\citep{calzavarini2005rayleigh}.

We do not observe a strong dependence on $Ra$ for
the rotation effects on the heat transfer and turbulent intensity. 
The curves of $Nu/Nu(0)$ and $Re/Re(0)$ measured for $Pr=10$ 
at $Ra =1.1 \times 10^7$ and $Ra = 2.2 \times 10^7$ are comparable
within the errorbars (see Fig.~\ref{fig2}). 
The only exception are the values of $Nu$ and $Re$ 
of the simulation at $Ra =1.1 \times 10^7$, $Ro=7.91\times 10^{-2}$. 
The inspection of the time serie of this simulation reveals that these
anomalous values are due to a single event of strong convection that influenced the whole statistics. 
In absence of rotation, the dependence of $Nu(0)$ and $Re(0)$ on $Ra$
is in agreement with the ultimate-state scaling laws 
$Nu(0) \propto Ra^{0.5}$ and $Re(0) \propto Ra^{0.5}$. 

%%%%%%%%%%%%%%%%%%%%%%
\begin{figure}
\centerline{\includegraphics[width=0.5\columnwidth]{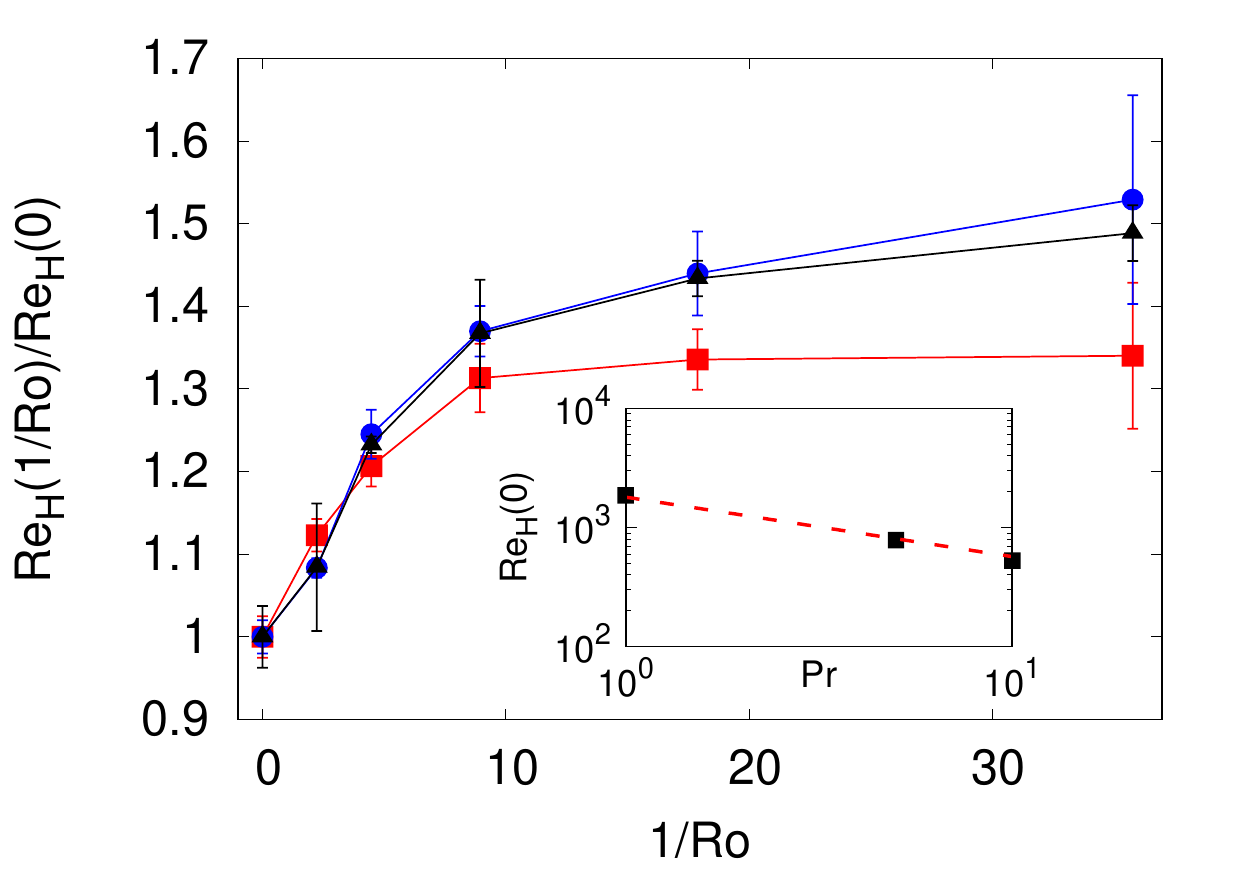}
\includegraphics[width=0.5\columnwidth]{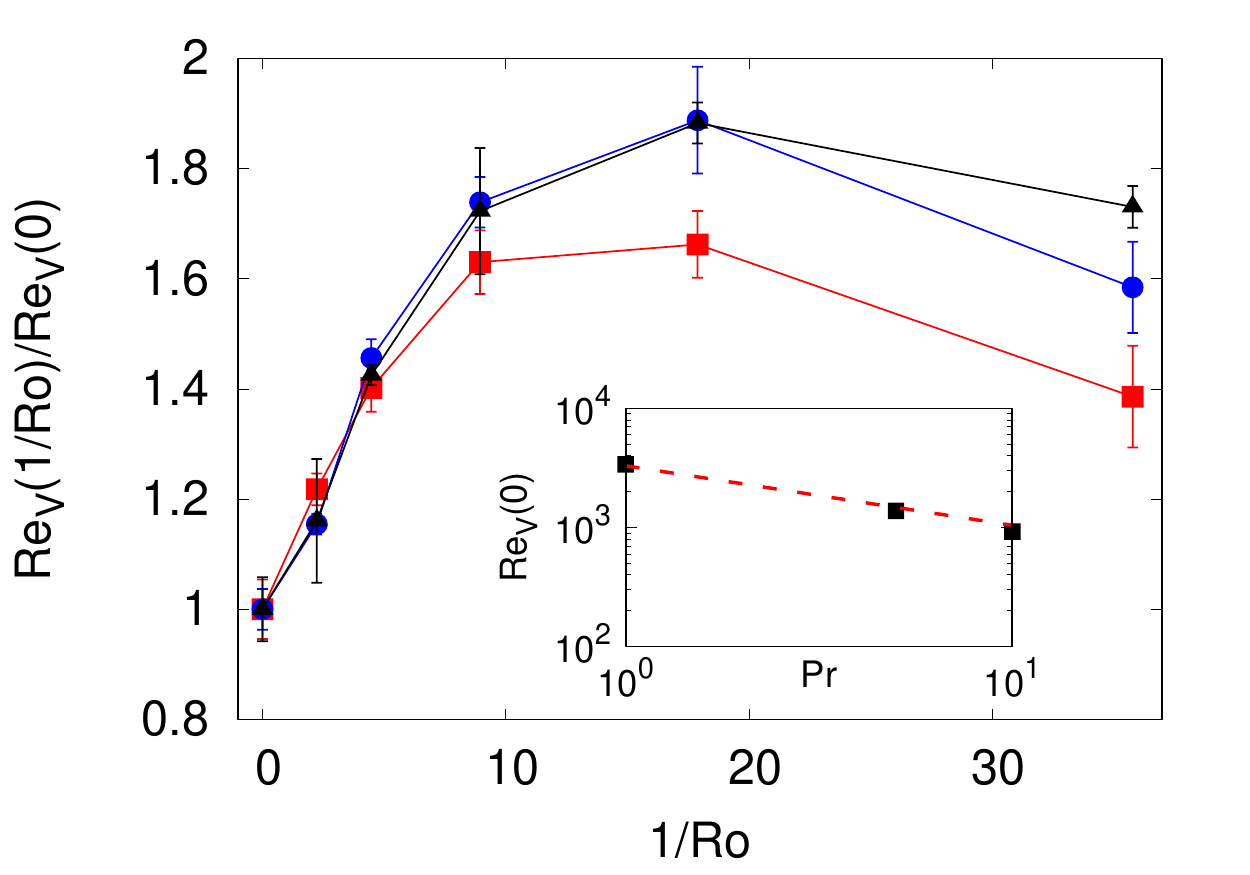}}
\centerline{\includegraphics[width=0.5\columnwidth]{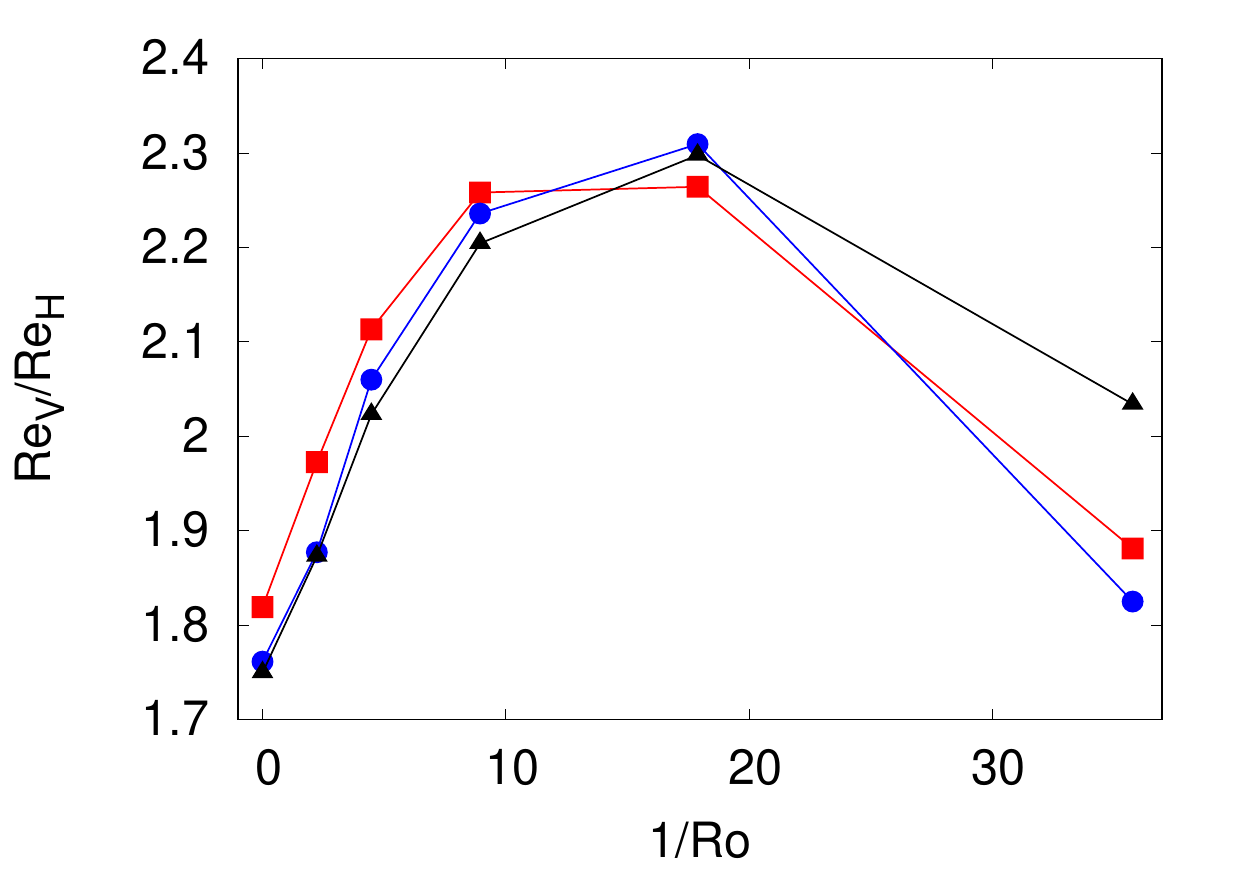}}
\caption{
Upper panels: $Re_H$ (a) and $Re_V$ (b) as a function
of $1/Ro$ normalized with the value at $1/Ro=0$ for simulations at 
$Ra=2.2 \times 10^7$ and 
$Pr=1$ (red squares), $Pr=5$ (blue circles) and $Pr=10$ 
(black triangles). The insets show the values of $Re_H$ and $Re_V$
in the absence of rotation ($1/Ro=0$) as a function of $Pr$. 
The dashed lines represent the scaling $Re_{H,V}(0) \propto Pr^{-1/2}$.
Lower panel: The ratio $Re_V/Re_H$ (c) as a function
of $1/Ro$ for simulations at $Ra=2.2 \times 10^7$ and 
$Pr=1$ (red squares), $Pr=5$ (blue circles) and $Pr=10$ 
(black triangles).
}
\label{fig:reynolds}
\end{figure}
%%%%%%%%%%%%%%%%%%%%%%
%%%%%%%%%%%%%%%%%%%%%%
\begin{figure}
\centerline{\includegraphics[width=0.5\columnwidth]{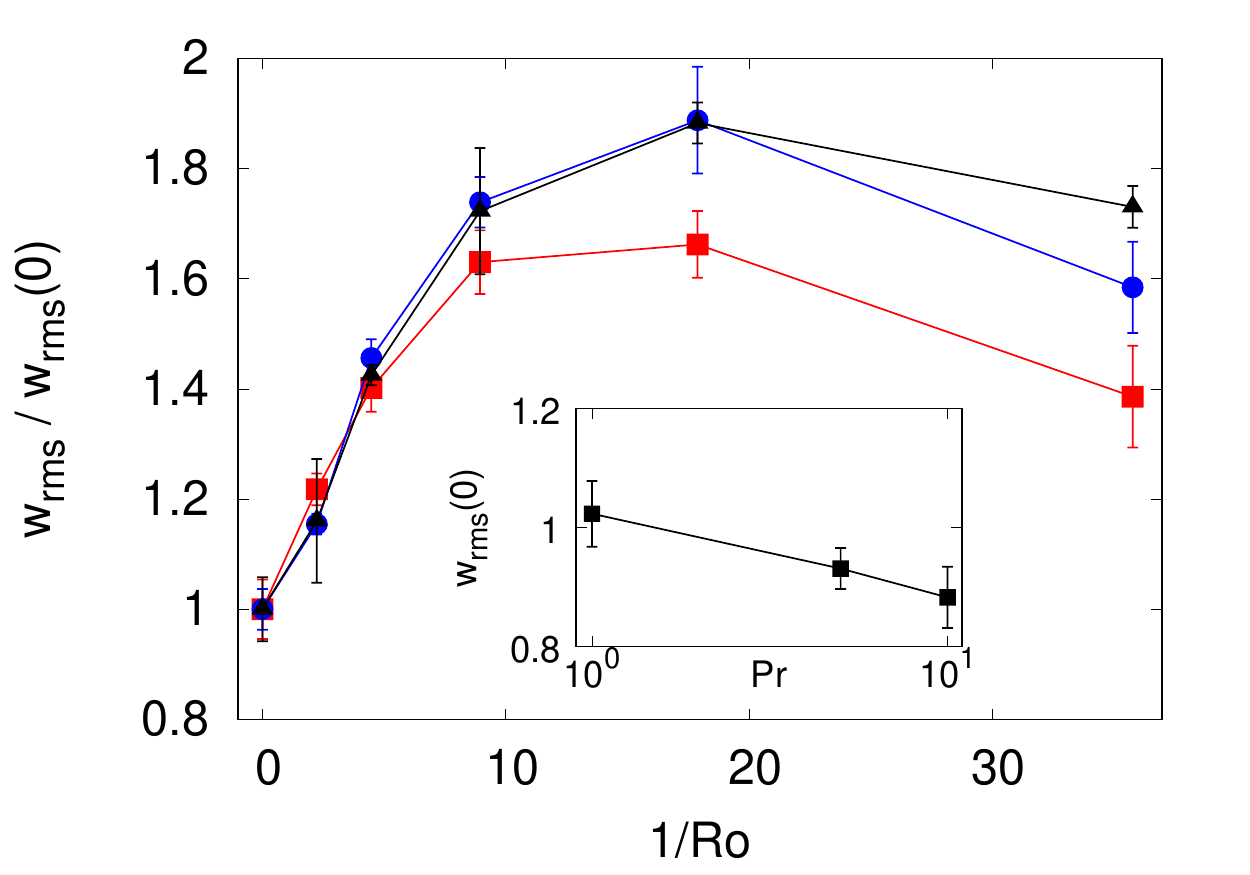}
\includegraphics[width=0.5\columnwidth]{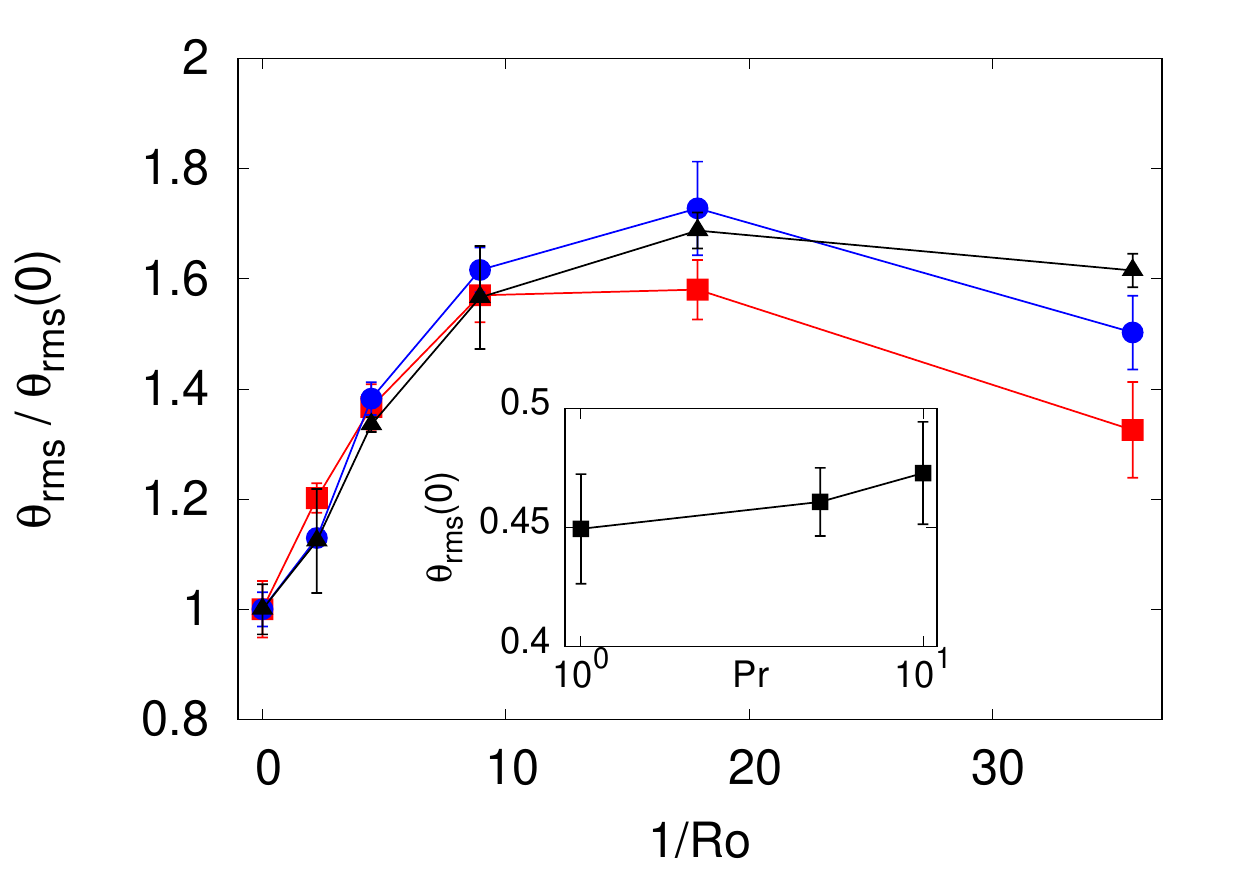}}
\centerline{\includegraphics[width=0.5\columnwidth]{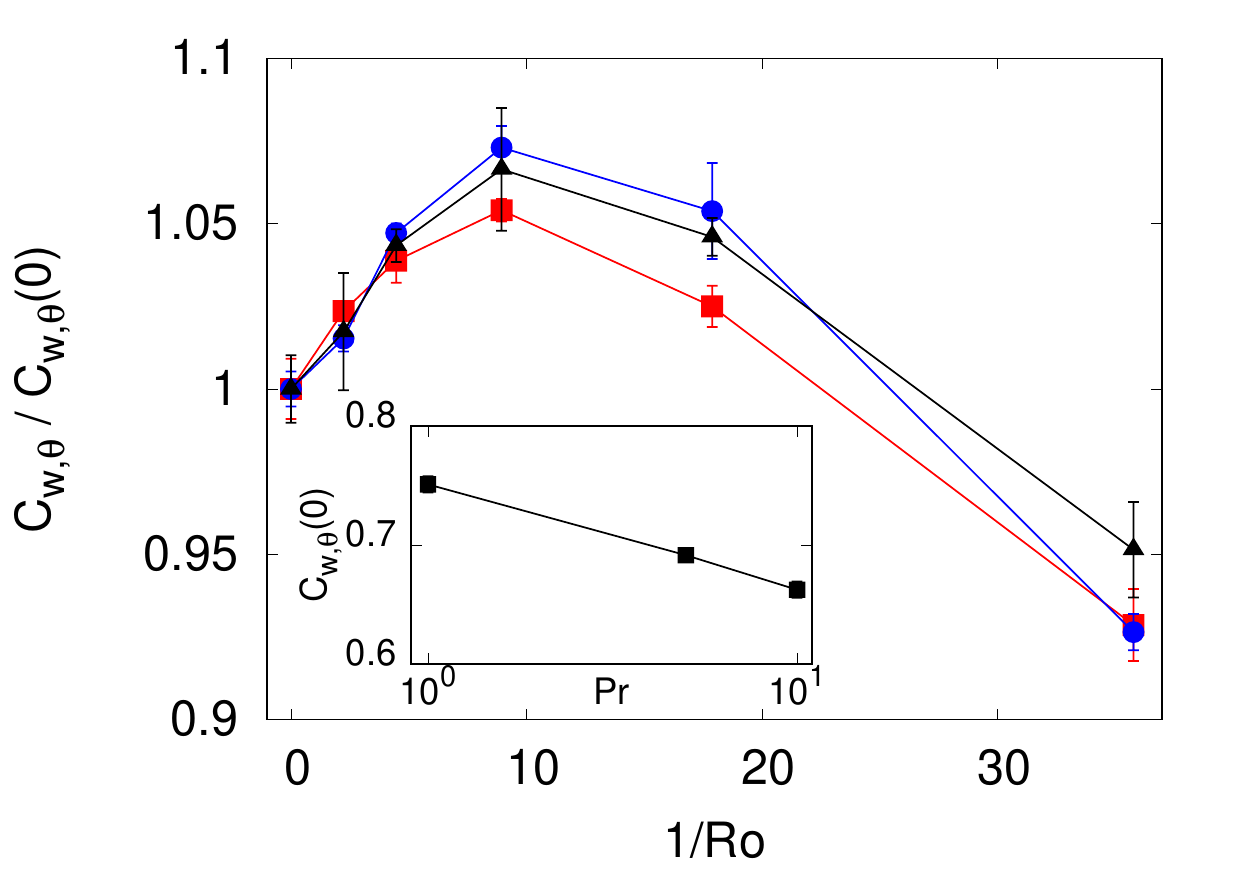}}
\caption{
Contributions to $Nu$: $w_{rms}$ (a), $\theta_{rms}$ (b) and $C_{w,\theta}$ (c) as a function
of $1/Ro$ normalized with their values at $1/Ro=0$ 
for simulations at $Ra=2.2 \times 10^7$ and
$Pr=1$ (red squares), $Pr=5$ (blue circles) and $Pr=10$
(black triangles).
The insets show the values of $w_{rms}(0)$, $\theta_{rms}(0)$ and
$C_{w,\theta}(0)$ in absence of rotation as a function of $Pr$.}
\label{fig:wtcor_rms}
\end{figure}
%%%%%%%%%%%%%%%%%%%%%%
The anisotropy between the horizontal and vertical
  velocity can be quantified by introducing the horizontal and vertical Reynold numbers defined respectively as:
\begin{equation}
Re_H = \frac{u_{rms} L}{\nu}\;,\;\;\;Re_V = \frac{w_{rms} L}{\nu}\;.
\label{eq:Rehv}
\end{equation}

In absence of rotation the dependence of $Re_H$ and $Re_V$ 
on $Pr$ is in agreement with the ultimate-state scalings
$Re_{H,V} \propto Pr^{-1/2}$ (see insets of Fig.~\ref{fig:reynolds}). 
The behavior of $Re_V$ as a function of $1/Ro$ is
non-monotonic and it is similar to the behavior of the total Reynolds
number, while the $Re_H$ shows a weaker monotonic increase.
In Fig.~\ref{fig:reynolds} we also show the ratio $Re_V/Re_H$ which
gives information on the anisotropy between vertical and horizontal
velocities.
The anisotropy, which is present already at $1/Ro=0$, is enhanced by
rotation and attains a maximum for $Ro \approx 6 \times 10^{-2}$.

Besides, following~\citet{boffetta2011effects} 
we decompose the Nusselt number as the product of three different contributions:  
\begin{equation}
Nu = \frac{w_{rms} \theta_{rms} C_{w,\theta}}
{\kappa \gamma}+1
\label{eq:Nusselt}
\end{equation}
where $C_{w,\theta}  = \langle w \theta \rangle / (w_{rms}
\theta_{rms}) $ 
is the correlation between the vertical velocity 
component $w$ and the temperature field $\theta$ .
All the three factors which contribute to $Nu$ display a
non-monotonic dependence on the rotation rate (see
Fig.~\ref{fig:wtcor_rms}). 
The largest variations are observed for the 
rms fluctuations of the vertical velocity and the
temperature, which for $Ro = 5.59\times10^{-2}$
are about $80\%$ larger than in the case $Ro=\infty$. 
The variation of the correlation $C_{w,\theta}$ is considerably
smaller. 

The dependence on $Pr$ of $w_{rms}$, $\theta_{rms}$, and
$C_{w,\theta}$ in absence of rotation (shown in the insets 
of  Fig.~\ref{fig:wtcor_rms}) has a simple physical interpretation. 
In order to increase $Pr$ keeping $Ra$ fixed, one has to 
increase the kinematic viscosity as $\nu \propto Pr^{1/2}$ 
and to decrease the thermal diffusivity as $\kappa \propto Pr^{-1/2}$. 
The increase of the viscosity suppresses the velocity fluctuations at
small scales, and therefore causes a decrease of $w_{rms}$. 
Conversely, the reduction of the thermal diffusivity allows
for the development of small-scale temperature fluctuations, 
and therefore causes an increase of $\theta_{rms}$. 
The opposite behavior of the small-scale structures of
the velocity and temperature fields at increasing $Pr$ 
causes the decrease of the correlation $C_{w,\theta}$. 

%%%%%%%%%%%%%%%%%%%%%%%%%%%%%%%%%%%%%%%%%%%%%%%%%%%%%%%%%%%%%%%%%%
\section{Columnar convective structures}
\label{sec4}

%%%%%%%%%%%%%%%%%%%%%%
\begin{figure}
\centerline{
\includegraphics[width=0.5\columnwidth]{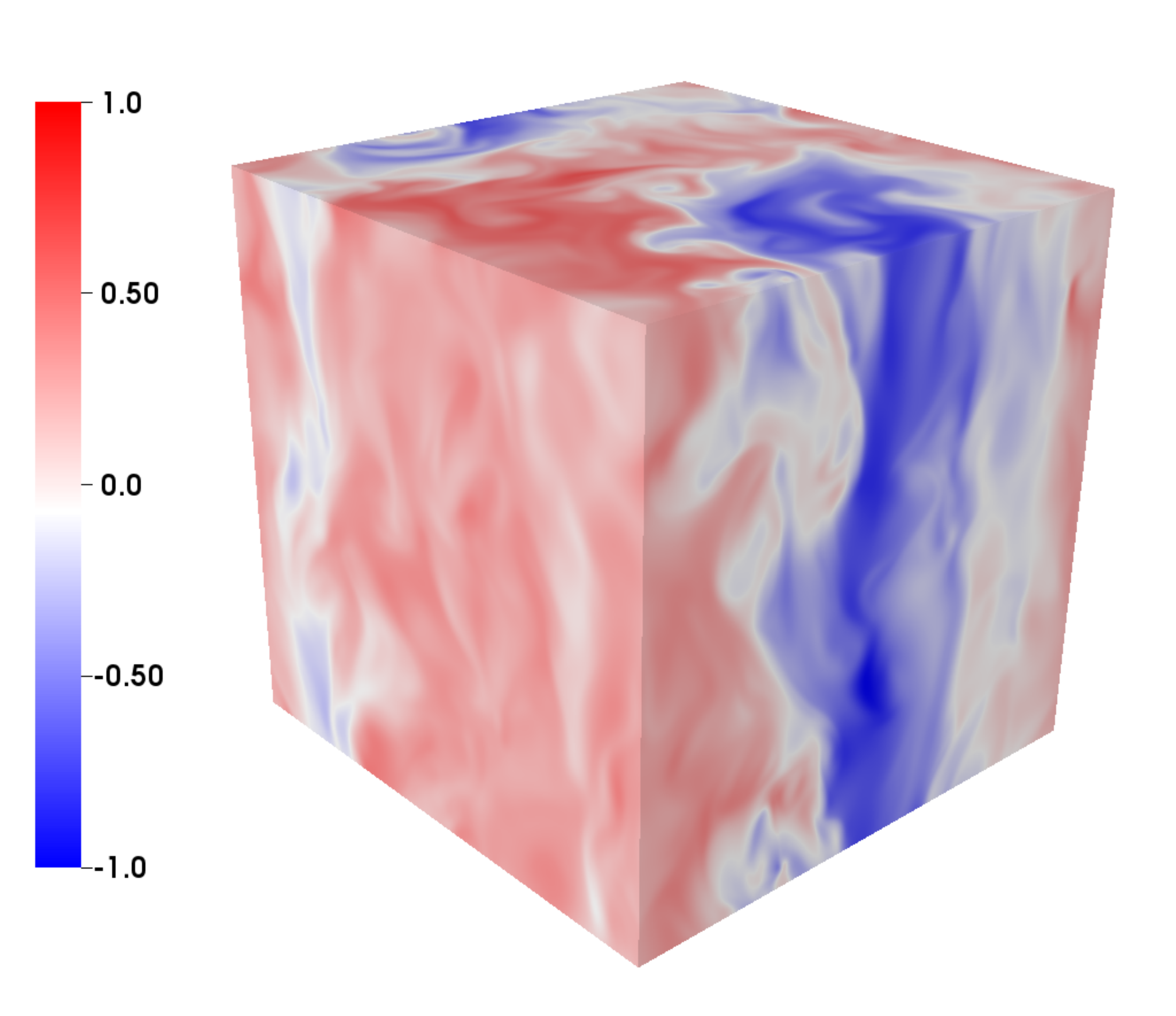}
\includegraphics[width=0.5\columnwidth]{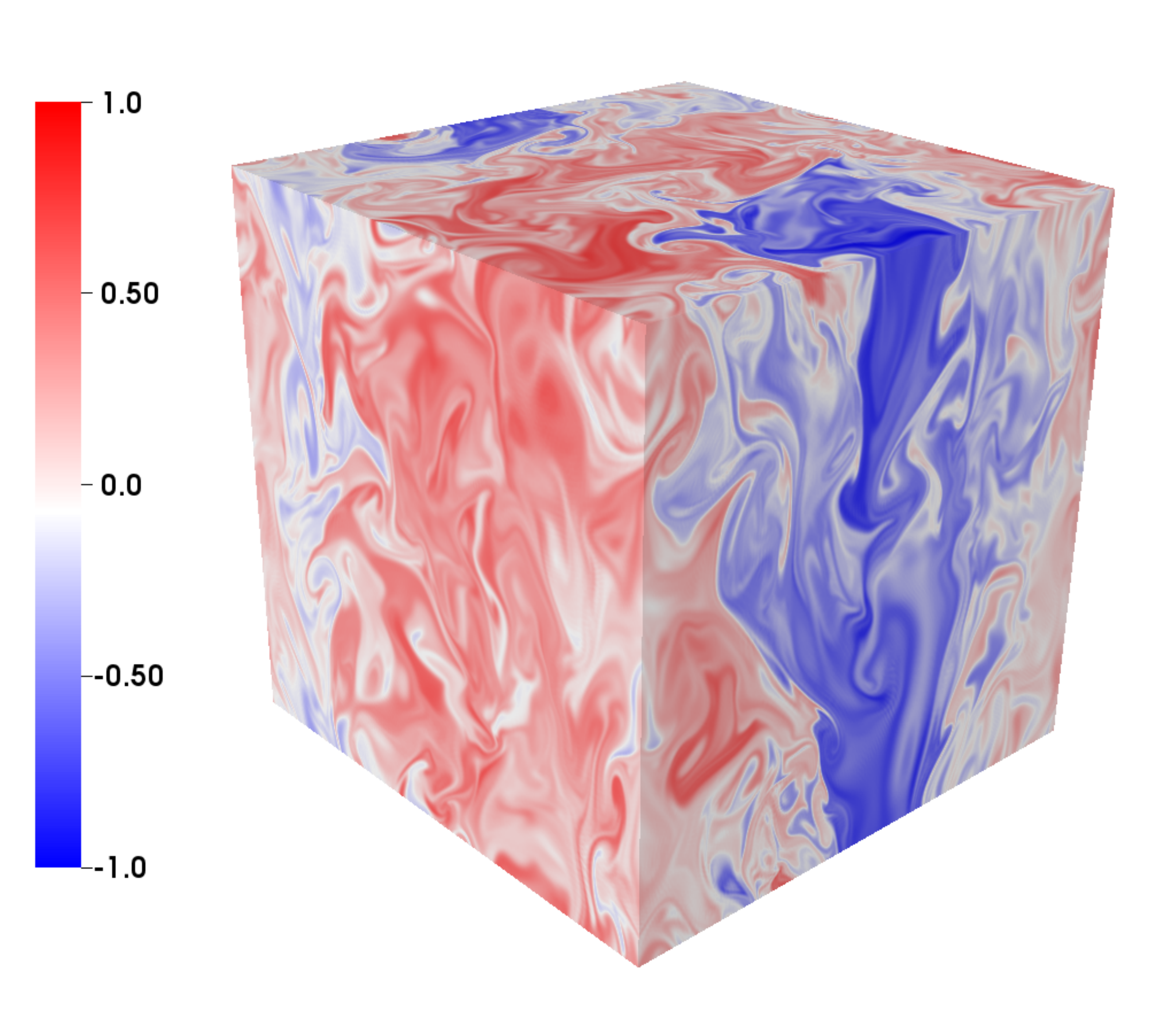}}
\centerline{
\includegraphics[width=0.5\columnwidth]{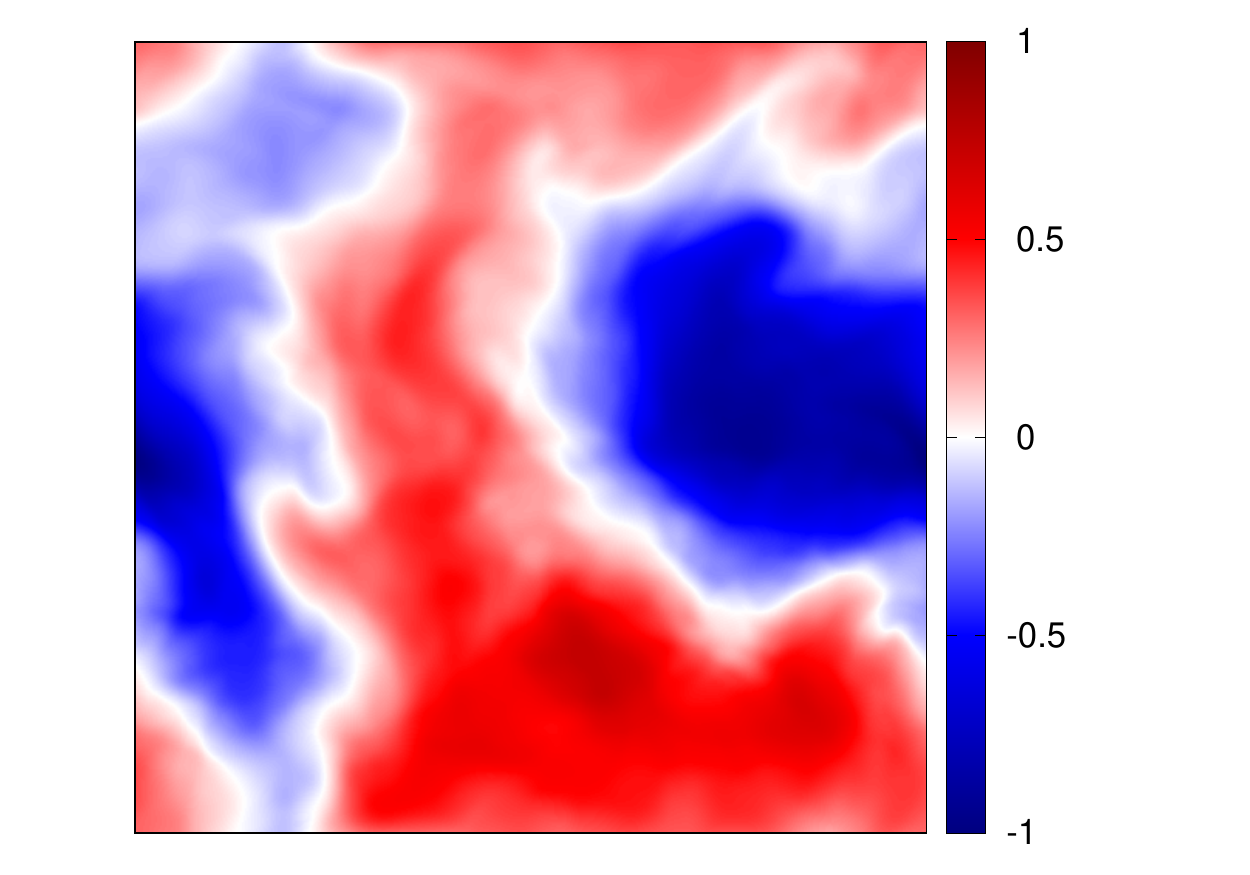}
\includegraphics[width=0.5\columnwidth]{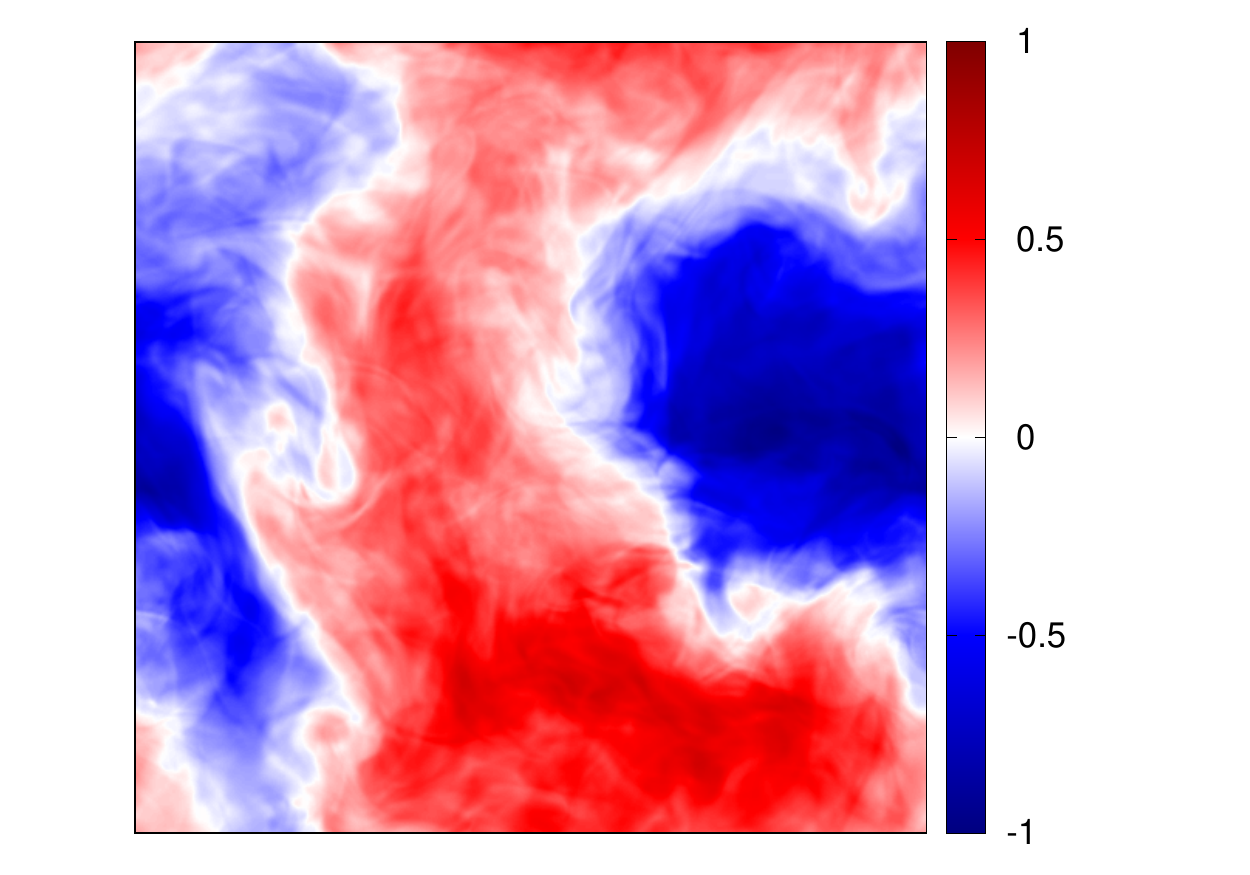}}
\caption{
Upper panels: 
Vertical velocity field $w$ (left panel) 
and temperature fluctuation field $\theta$ (right panel)
during a strong convective event at time $t=80\tau$ 
in the simulation with $Ra = 2.2 \times 10^7$, $Pr=10$ and $Ro= 5.59 \times 10^{-2}$. 
Lower panels: 
Two-dimensional fields $w^{2D}$ (left) and $\theta^{2D}$ (right) 
obtained by averaging the fields $w$ and $\theta$ shown above
along the vertical direction.
Fields are rescaled with maxima of absolute values.}
\label{fig:3D2Dfields}
\end{figure}
%%%%%%%%%%%%%%%%%%%%%%

The time series of the Nusselt number obtained in our simulations 
are characterized by strong fluctuations, which correspond to events of weak/strong convection. 
The standard deviation of these fluctuations 
is of the order of $50\%$ of their mean values, 
defined as the time-average over the duration of the simulations 
(and corresponding to the values reported in the previous section). 

We have found that, in the rotating cases, the events of strong
convection are related with the formation of columnar structures
aligned with the rotation axis, which are present both in the
temperature field and in the vertical velocity field. 
As an example, we show in Figure~\ref{fig:3D2Dfields} 
the field $\theta$ and $w$ at time $t=80\tau$,
corresponding to a local maximum of the time series of $Nu$ 
in the simulation with $Ra = 2.2 \times 10^7$, $Pr=10$ and $Ro= 5.59
\times 10^{-2}$.

The presence of quasi-2D columnar structures 
is a distinctive feature of rotating turbulence, 
and has been observed both in
experiments~\citep{hopfinger1982turbulence,staplehurst2008structure,gallet2014scale} and numerical
simulations~\citep{yeung1998numerical,yoshimatsu2011columnar,biferale2016coherent}. 
The formation of columnar structures has been reported also on the
case of RB convection by~\cite{kunnen2010vortex}. 
In the case of BTC we observe a significant correlation between hot (cold) regions 
and rising (falling) regions in the core of these structures, which
results in a strong increase of the heat flux. 

\begin{figure}
\centerline{
\includegraphics[width=0.5\columnwidth]{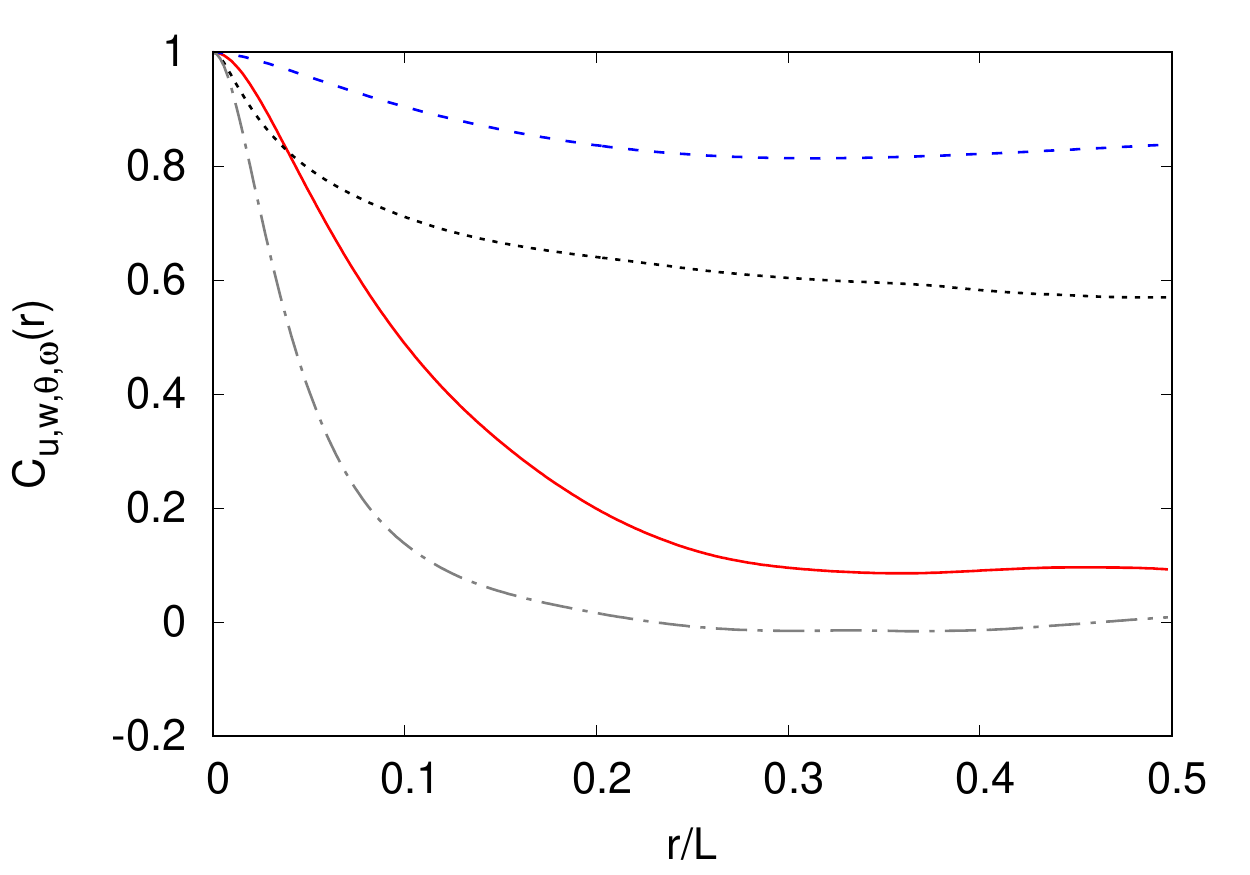}
\includegraphics[width=0.5\columnwidth]{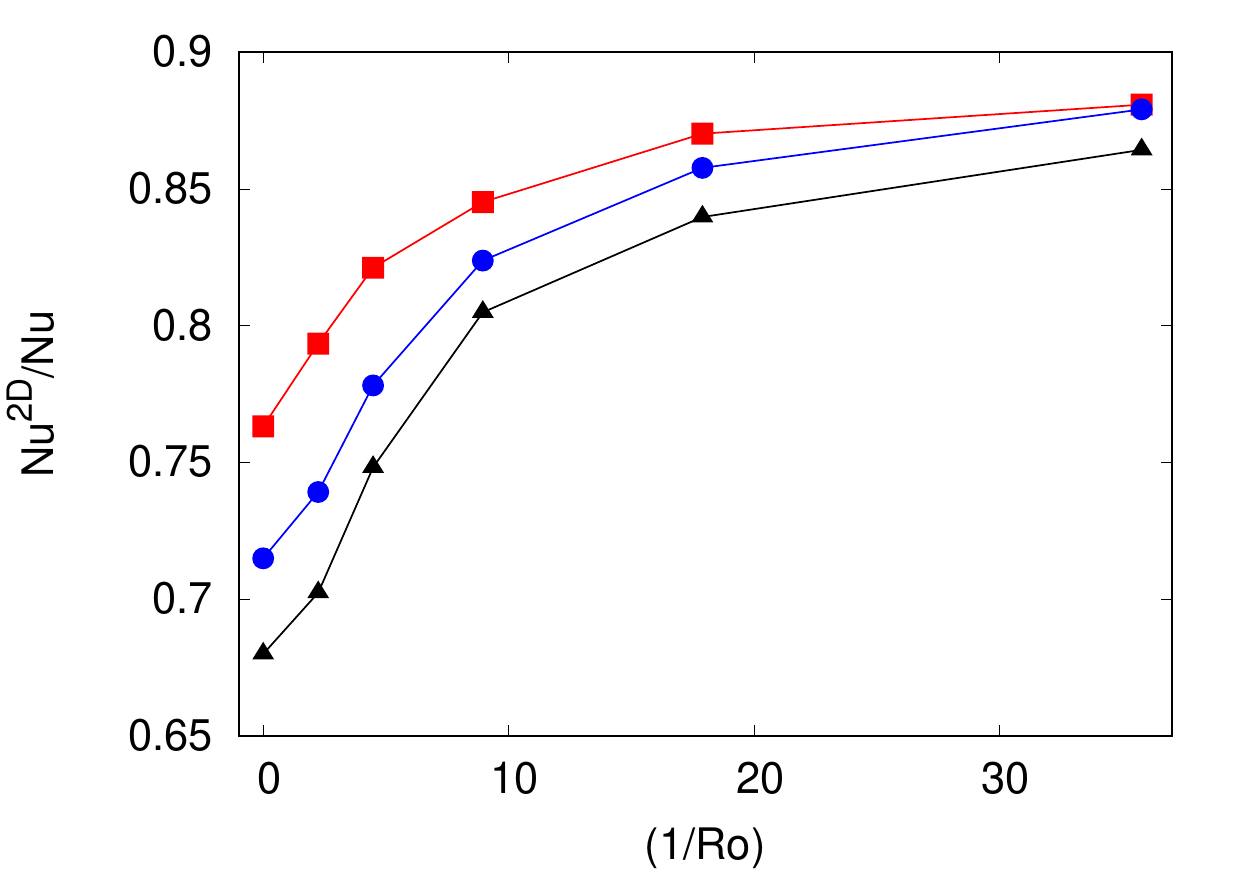}}
\caption{
Left panel: 
Correlation function of horizontal velocity $C_{u}(r)$ (red line),
vertical velocity $C_{w}(r)$ (blue dashed line), temperature
$C_{\theta}(r)$ (black dotted line) and vertical vorticity $C_{\omega_z}(r)$ (grey dash-dotted line)
at time  $t=80\tau$ for $Ra = 2.2 \times 10^7$, $Pr=10$ and $Ro= 5.59
\times 10^{-2}$.  
Right panel: 
Ratio $Nu^{2D}/Nu$ as a function of $1/Ro$ for $Ra = 2.2 \times 10^7$,
$Pr=1$ (red squares), $Pr=5$ (blue circles) and $Pr=10$ (black triangles) 
}
\label{fig:bidim}
\end{figure}

In order to investigate quantitatively this phenomenon we proceed as follow. 
First, we measure the degree of bidimensionalization of the system
during an event of strong convection, 
by studying how much the velocity and temperature fields (at fixed time) are correlated in the vertical direction. 
For this purpose, we 
computed the vertical correlation function of $u$, $w$, $\theta$ and the z-component of the vorticity $\omega_z$: 
\begin{equation}
 C_{u}(r) = \langle u({\bf x}+r {\bf \hat{e}_3}) \; u({\bf x}) \rangle  
 \label{eq8}
\end{equation}\begin{equation}
 C_{w}(r) = \langle w({\bf x}+r {\bf \hat{e}_3}) \; w({\bf x}) \rangle  
 \label{eq9}
\end{equation}
\begin{equation}
 C_{\theta}(r) = \langle \theta({\bf x}+r {\bf \hat{e}_3}) \; \theta({\bf x}) \rangle  
 \label{eq10}
\end{equation}
\begin{equation}
 C_{\omega_z}(r) = \langle \omega_z({\bf x}+r {\bf \hat{e}_3}) \; \omega_z({\bf x}) \rangle  
 \label{eq10}
\end{equation}
In Fig.~\ref{fig:bidim} we show a comparison of the vertical
correlation functions computed in the case of the simulation with $Ra = 2.2 \times 10^7$, $Pr=10$ and $Ro= 5.59
\times 10^{-2}$ at the same time of the Figure~\ref{fig:3D2Dfields}
($t=80\tau$). 
At variance with the typical columnar vortices
  observed in rotating turbulence, here we do not find a strong
  vertical correlation of the z-component of the vorticity (see
  Fig.~\ref{fig:bidim}).
Also the vertical correlation of horizontal velocity $u$ 
decays at scales larger than $1/2$ of the box size. Conversely the vertical
velocity $w$ and the temperature fields $\theta$ remains correlated
through the whole domain.

%%%%%%%%%%%%%%%%%%%%%%
%%%%%%%%%%%%%%%%%%%%%%

This long-scale, vertical correlation lead us to introduce the $2D$ fields
$w^{2D}=\langle w \rangle_z$ and $\theta^{2D}= \langle
\theta \rangle_z$, defined as the average along the vertical direction 
of the respective $3D$ fields.
In Fig.~\ref{fig:3D2Dfields} (lower panels) we show the $2D$ fields 
of $w^{2D}$ and $\theta^{2D}$ obtained for the simulation at 
$Ra = 2.2 \times 10^7$, $Pr=10$ and $Ro= 5.59
\times 10^{-2}$ at time $t=80\tau$, which confirms the spatial correlation between the hot (cold) regions and the
rising (falling) regions also in the vertically averaged fields. 

Despite the lack of a strong vertical
  correlation of $\omega_z$, the inspection of the 2D field
  $\omega_z^{2D}=\langle \omega_z \rangle_z$
reveals a connection between the regions of intense heat flux, which
can be identified as thermal convective columns, and cyclonic regions,
i.e. those which rotates in the same direction of $\boldsymbol\Omega$. It is
possible that the preferential link between convective structures and cyclones
could be related with the cyclonic-anticyclonic asymmetry which is
observed in rotating turbulence (for a recent review on rotating
turbulence see \citet{Godeferd2015}). 

Finally, we introduce the $2D$ Nusselt number defined in terms of the $2D$ fields
as : 
\begin{equation}
  Nu^{2D}= \langle \frac{\langle w \rangle_z \langle \theta \rangle_z}{\kappa\gamma} \rangle_{x,y}
  \label{eq:Nu2D}
\end{equation}
where $\langle \cdots \rangle_{x,y}$ is the average over the
horizontal directions $x$ and $y$. 
The physical meaning of the ratio $Nu^{2D}/Nu$ 
is the relative contribution of the 2D modes, i.e. of the columnar
structures, to the total heat
transport. In Figure~\ref{fig:bidim} we show the ratio $Nu^{2D}/Nu$ for the various $Pr$ and $Ro$ simulations 
at $Ra=2.2 \times 10^7$. The increase of $Nu^{2D}/Nu$ with the
rotation rate demonstrate that in the limit of vanishing $Ro$ 
the heat transport is dominated by the 2D modes. 
We also observe a systematic trend as a function of $Pr$: increasing
$Pr$ reduces the contribution of the 2D modes to the heat flux. 
This effect can be understood in terms of the reduced spatial correlation
between the fields $w$ and $\theta$ at increasing $Pr$, as discussed
in the previous Section (see Fig.~\ref{fig:wtcor_rms} and the related discussion).  

%%%%%%%%%%%%%%%%%%%%%%%%%%%%%%%%%%%%%%%%%%%%%%%%%%%%%%%%%%%%%%%%%
\section{Conclusions}
\label{sec5}
We have investigated the behavior of the bulk turbulence convection (BTC) 
system in a rotating frame by performing extensive
DNS of the Boussinesq equations for an incompressible flow in 
a cubic box with periodic boundary conditions in all directions.
In the absence of rotation, we confirmed the consistency of the both 
$Nu$ and $Re$ scaling with $Pr$ and $Ra$ numbers according to the 
``ultimate regime of thermal convection'' theory \citep{Grossmann2000}. 
In the presence of rotation, quantified by the Rossby number $Ro$,
we find a surprising strong enhancement of both $Nu$ and $Re$ for 
intermediate values of $Ro$ followed by a moderate decreases for the 
largest $Ro$ investigated. 

A detailed analysis of the temperature and velocity fields shows 
that the observed heat flux enhancement at intermediate rotation is 
due to the formation of columnar convective structures with strong 
correlations between temperature and vertical velocity. 

The understandig of the mechanism behind this phenomenom is still incomplete.
In the RB case the non-monotonic increase of $Nu$ is associated with
the Ekman pumping and it depends on the modification of the boundary layer
caused by rotation.
Even if in BTC case the boundary layer is absent we still observe
similarities with RB phenomenology. In particular we find a correlation between
$Nu$ and vertical velocity variations.
Further studies are required in order to improve our knowledge on this
phenomenon.

%%%%%%%%%%%%%%%%%%%%%%%%%%%%%%%%%%%%%%%%%%%%%%%%%%%%%%%%%%%%%%%%%
\section{Acknowledgments}
\label{sec6}
We thank Detlef Lohse for stimulating discussions.
We acknowledge support by the Departments of Excellence grant (MIUR).  Moreover
HPC  Center  CINECA  is  gratefully  acknowledged  for  computing  resources
(INFN-Cineca grant No. INF18-fldturb). 

%%%%%%%%%%%%%%%%%%%%%%%%%%%%%%%%%%%%%%%%%%%%%%%%%%%%%%%%%%%%%%%%%
\bibliographystyle{jfm}
\bibliography{biblio}

\end{document}